\documentclass[12pt,preprint]{aastex}

\def\ltsima{$\; \buildrel < \over \sim \;$}
\def\gtsima{$\; \buildrel > \over \sim \;$}
\def\lsim{\lower.5ex\hbox{\ltsima}}
\def\gsim{\lower.5ex\hbox{\gtsima}}
\def\lapp{\ifmmode\stackrel{<}{_{\sim}}\else$\stackrel{<}{_{\sim}}$\fi}
\def\gapp{\ifmmode\stackrel{>}{_{\sim}}\else$\stackrel{<}{_{\sim}}$\fi}

\def\msol{\,\mathrm{M}_\odot}

\newcommand{\masyr}{${\rm mas\, yr^{-1}}$}

\newcommand{\masyrmag}{${\rm mas\, yr^{-1}\, mag^{-1}}$}
\newcommand{\gaia}{{\it Gaia~}}
\newcommand{\kms}{${\rm km\, s^{-1}}$}
\newcommand{\kmsdeg}{${\rm km\, s^{-1}\, deg^{-1}}$}
\newcommand{\at}{\makeatletter @\makeatother}

\usepackage{amsmath}
\usepackage{textcomp}
\usepackage{amssymb}
\usepackage{multirow}
\usepackage{booktabs}
\usepackage{graphicx}
\usepackage[export]{adjustbox}
\usepackage{subfigure}

\newdimen\minuswidth    
\setbox0=\hbox{$-$}
\minuswidth=\wd0
\catcode`@=\active
\def@{\kern\minuswidth}
\setbox0=\hbox{\rm0}

\shorttitle{}
\shortauthors{Massari et al.}

\begin{document}
\title{3D motions in the Sculptor dwarf galaxy as a glimpse of a new era}

\author{{\rm
D. Massari\altaffilmark{1,2,\ast},
M. A. Breddels\altaffilmark{1},
A. Helmi\altaffilmark{1},
L. Posti\altaffilmark{1},
A. G. A. Brown\altaffilmark{2},
E. Tolstoy\altaffilmark{1}
}}

\affil{\altaffilmark{1}{\rm \it \small Kapteyn Astronomical Institute, University of Groningen,
PO Box 800, 9700 AV Groningen, The Netherlands}}
\affil{\altaffilmark{2}{\rm \it  \small Leiden Observatory, Leiden University, P.O. Box 9513, 2300 RA Leiden, The Netherlands}}

\altaffiltext{$^\ast$}{Corresponding author: massariATastro.rug.nl}

{\bf The 3D motions of stars in small galaxies beyond our own are
  minute and yet they are crucial for our understanding of the nature
  of gravity and dark matter$^{1,2}$. Even for the
  dwarf galaxy Sculptor which is one of the best studied systems and
  inferred to be strongly dark matter
  dominated$^{3,4}$, there are conflicting
  reports$^{5,6,7}$ on its mean motion around the
  Milky Way and the 3D internal motions of its stars have never been
  measured.  Here we report, based on data from the {\it Gaia} space
  mission$^8$ and the {\it Hubble Space Telescope}, a new
  precise measurement of Sculptor's mean proper motion. From this we
  deduce that Sculptor is currently at its closest approach to the
  Milky Way and moving on an elongated high-inclination orbit that
  takes it much farther away than previously thought. For the first
  time we are also able to measure the internal motions of stars in
  Sculptor.  We find $\sigma_{R}=11.5 \pm 4.3$~\kms~ and
  $\sigma_{T}=8.5\pm3.2$~\kms~ along the projected radial and
  tangential directions, implying that the stars in our sample move
  preferentially on radial orbits as quantified by the anisotropy
  parameter, which we find to be $\beta\sim 0.86^{+0.12}_{-0.83}$ at a
  location beyond the core radius. Taken at face value such a high
  radial anisotropy requires abandoning conventional
  models$^9$ for the mass distribution in Sculptor. Our
  sample is dominated by metal-rich stars and for these
  we find $\beta^{MR} \sim 0.95^{+0.04}_{-0.27}$, a value consistent
  with multi-component models where Sculptor is embedded in a cuspy
  dark halo$^{10}$ as expected for cold dark
  matter.}

\begin{figure}
 \centering%
\includegraphics[scale=0.5]{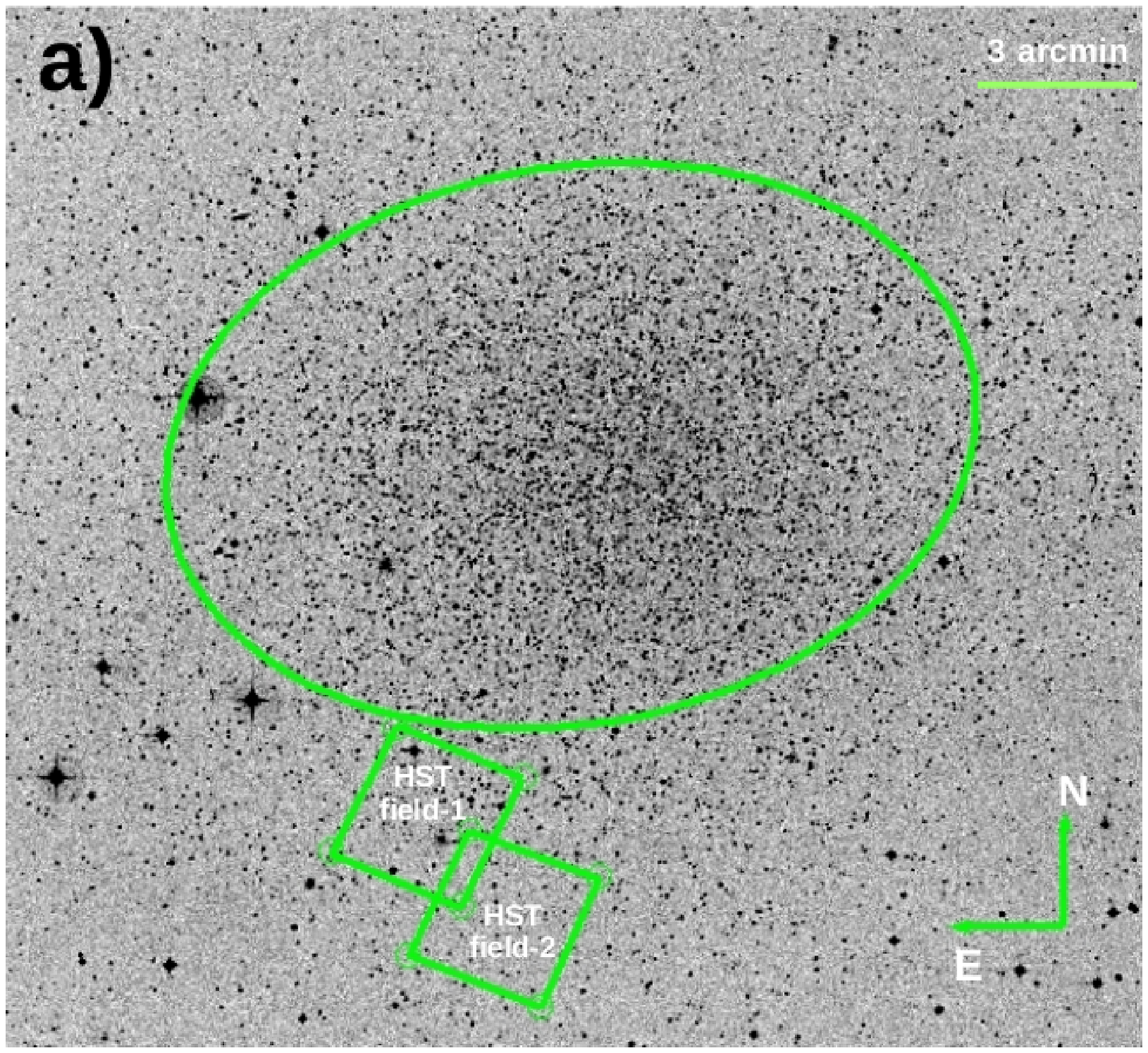}\\
\includegraphics[scale=0.2]{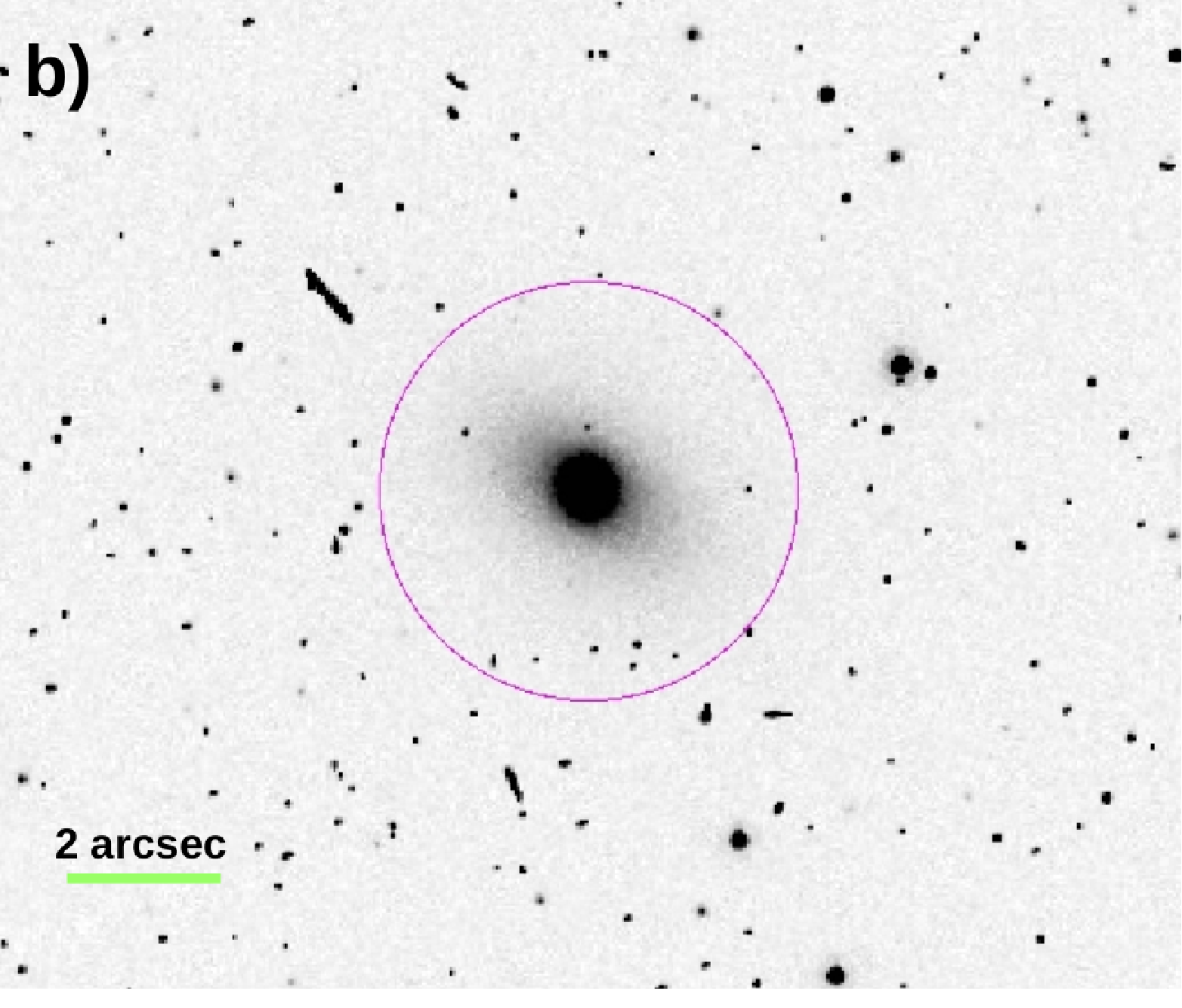} 
\includegraphics[scale=0.2]{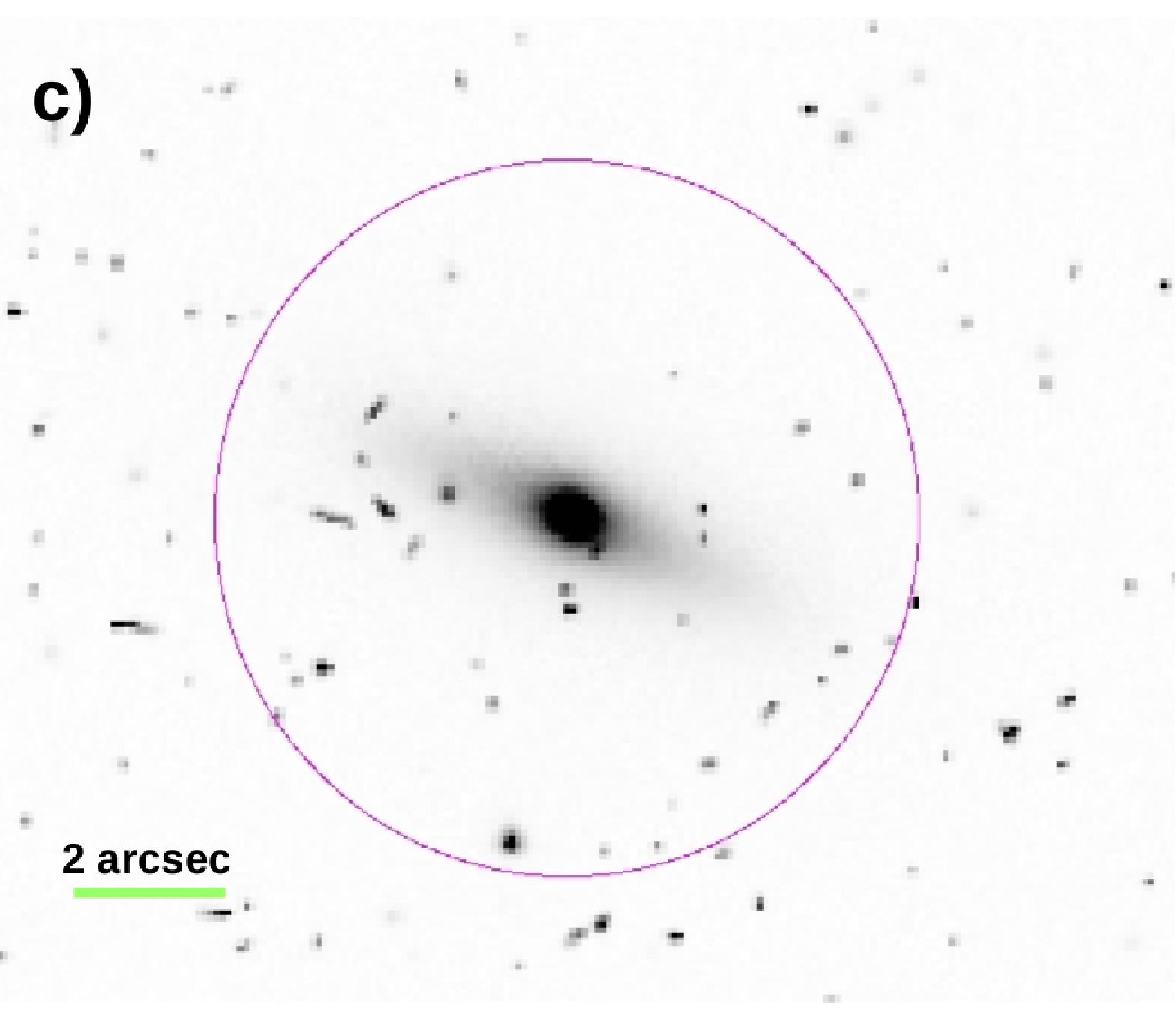}
\caption{{\bf Field of view towards the Sculptor dwarf spheroidal galaxy.} a) is a Digital Sky Survey 
  image of the center of Sculptor. The ellipse indicates the 
  core radius$^{13}$ ($r_c \sim 5.9$\arcmin $\sim144$ pc). The two HST pointings marked with
  boxes are located at an
  average distance $R_{HST} \sim 7.6\arcmin$ $\sim185$ pc, well inside the half-light
  radius (r$_{hl}\sim 16\arcmin \sim 390$ pc) of the system. b) and c) show the HST images of the two background galaxies
  used to determine the absolute zero point of the PM.}\label{fov}
\end{figure}

To measure the proper motions (PMs) of individual stars in Sculptor we used data taken
$12.27$ years apart. The first epoch was acquired with 
the Advanced Camera for Surveys on board
HST.  The data set consists of two
overlapping pointings separated by about 2\arcmin~($\sim50$ pc, see
Fig.~\ref{fov}), each split in several 400 sec exposures in the F775W filter. The overlapping field-of-view has 
been observed 11 times. We obtained a catalog of positions, instrumental
magnitudes and Point Spread Function (PSF) fitting-quality parameters
by treating each chip of each exposure
independently. Stellar positions were corrected for
filter-dependent geometric distortions$^{11}$.  We then cross-matched the single catalogs to compute
3$\sigma$-clipped average positions, magnitudes and corresponding
uncertainties. 
We built the complete HST catalog after excluding all
the saturated sources and those that were measured less than 4
times. The second epoch is provided by
the \gaia first data release$^{12}$.  We extracted from the \gaia archive all sources in the direction of Sculptor.

\begin{figure*}[!t]
 \centering%
\includegraphics[scale=0.26]{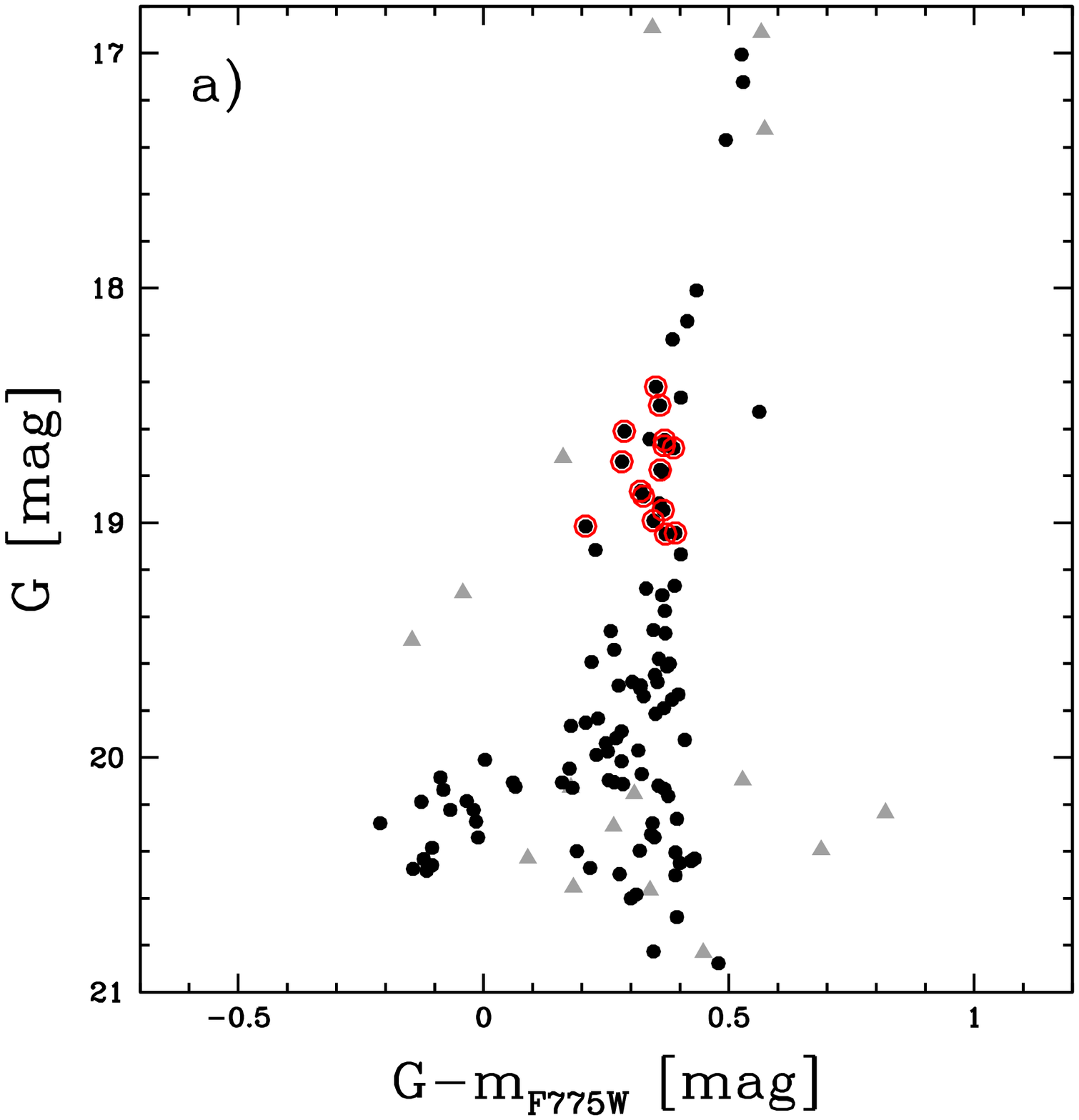}
\includegraphics[scale=0.26]{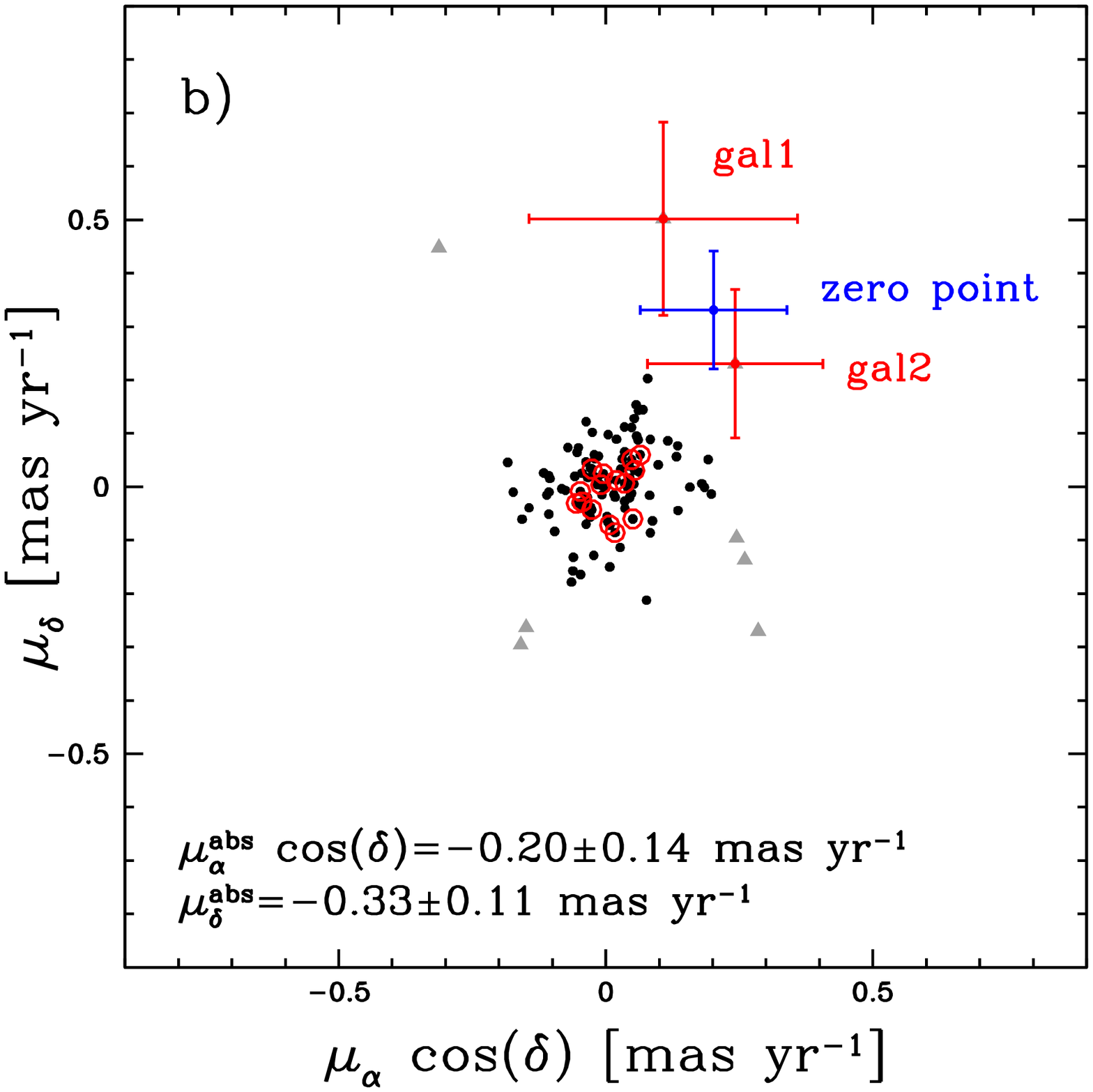}
\includegraphics[scale=0.26]{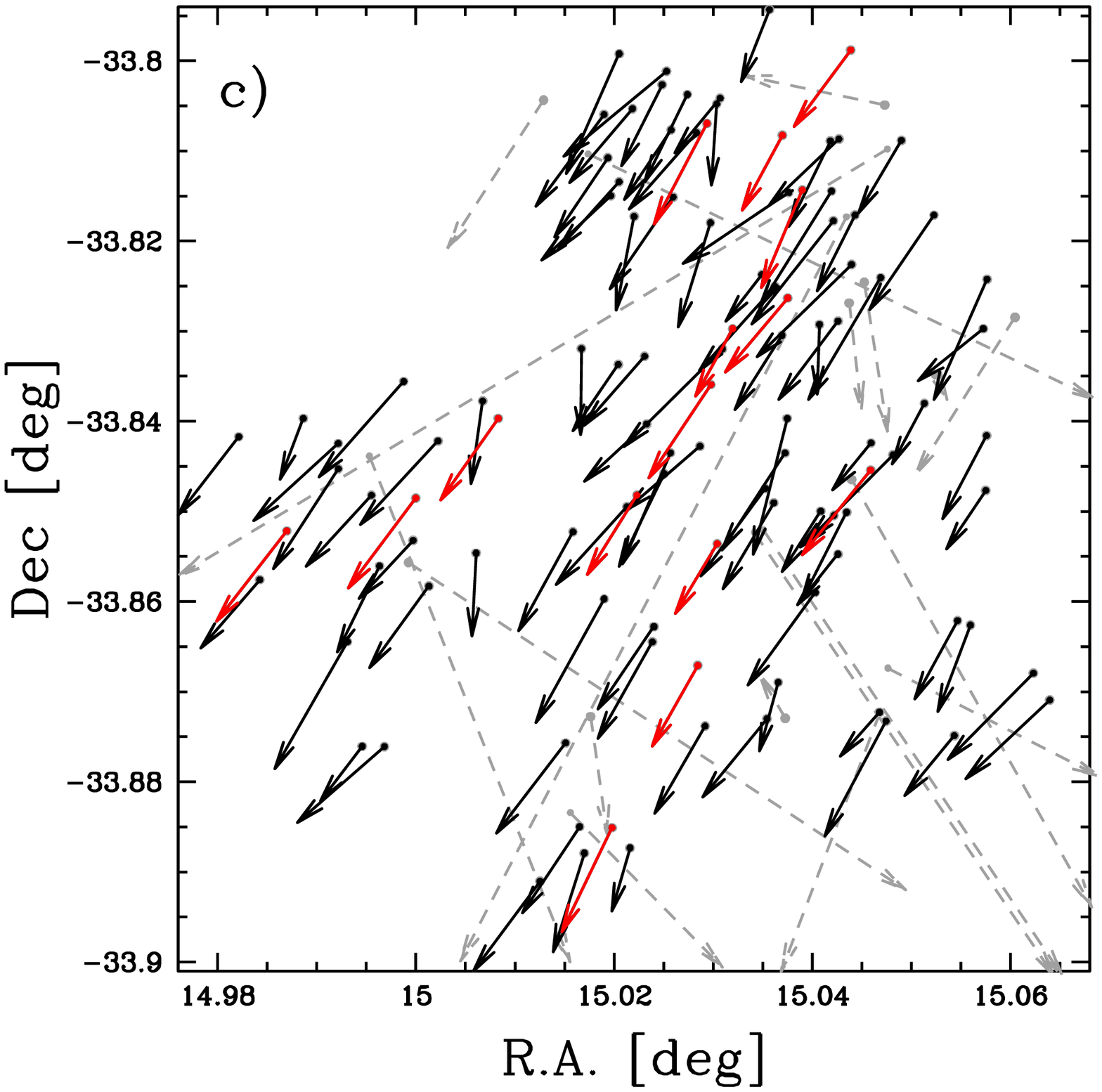}
\caption{{\bf Properties of our sample.} a) is the color-magnitude diagram for the stars in our PM
  catalog. Black dots are likely members (with PM amplitude smaller
  than 0.23 \masyr), red circles are the 15 member stars with the best
  measured PMs (used to compute the internal velocity dispersion of
  Sculptor), and gray triangles are likely non-members. The same color 
  coding is used in the next panels. b) shows 
  the sources with a measured PM. The two background galaxies are
  marked in red, and their weighted mean in blue, together with the associated 1$\sigma$ uncertainty. c) shows the
  observed projected motions of stars in the field.}\label{vpd}
\end{figure*}

We transformed the HST positions to the equatorial reference frame
defined by the \gaia data (right ascension, RA, and declination, DEC), using a six-parameter linear
transformation$^{14}$. We found $126$ stars in common and their PMs were computed as the
difference between the \gaia and HST positions, divided by the
temporal baseline.  The uncertainties on the PMs were computed as the
sum in quadrature between the \gaia and HST positional errors,
divided by the temporal baseline, also taking into account the
non-negligible correlations between {\it Gaia}'s RA and DEC
uncertainties.
After this first iteration, we repeated the procedure several times to
compute the frame transformations using only likely
members of Sculptor. These were selected using their location
in the (G, G-m$_{F775W}$) color-magnitude diagram (Fig.~\ref{vpd}a)
and their previous PM determination.  
After three iterations, the number of selected stars
stabilized at $91$.  

\begin{figure}
 \centering%
\includegraphics[scale=0.5]{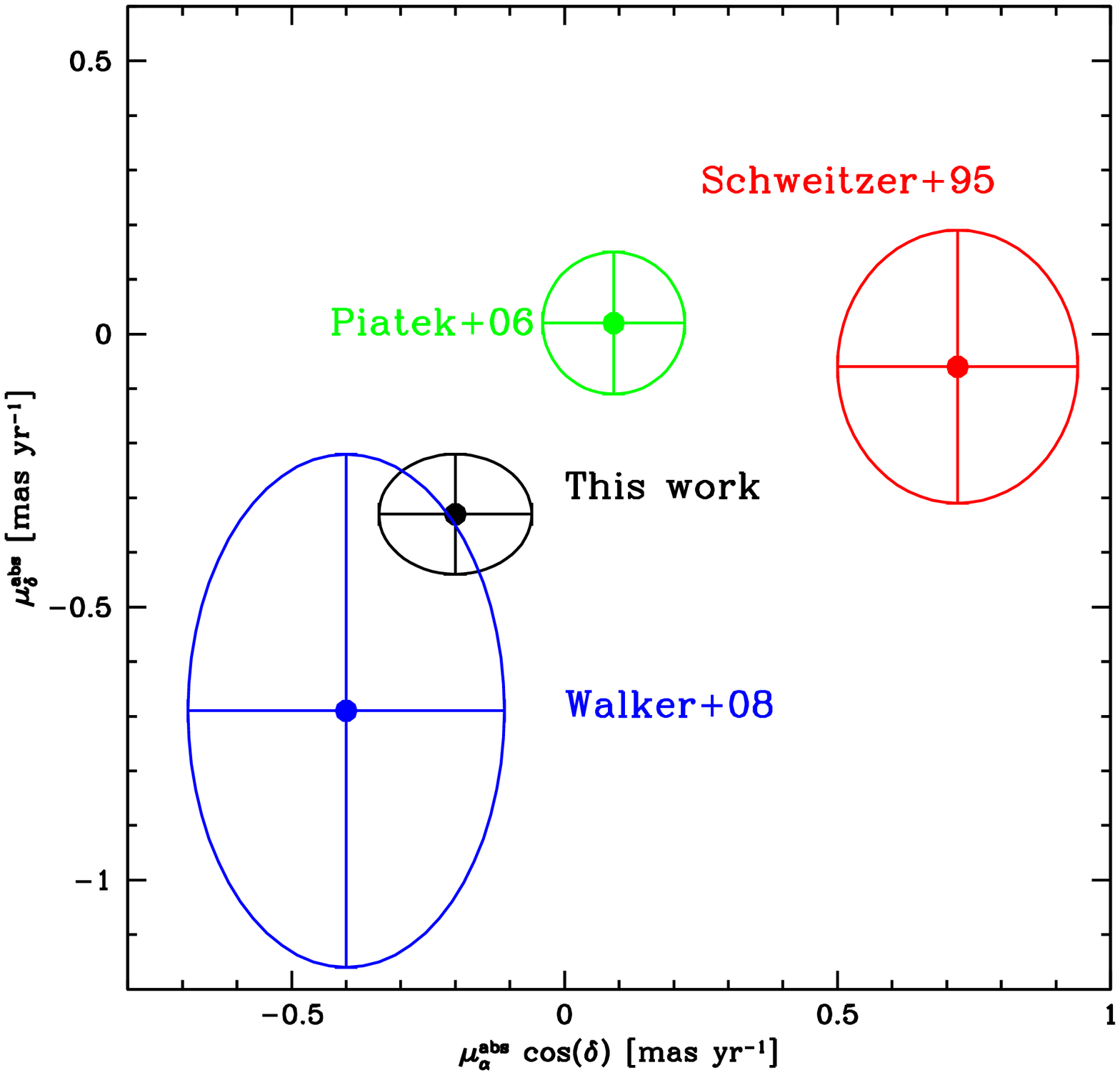}
\caption{{\bf Comparison to previously published PM estimates for
  Sculptor}. Each
  ellipse denotes the $68$\% confidence level. 
  It is not very surprising that none of the PMs agree with
  each other at this level as the two
  astrometric measurements are based either on photographic plates$^{5}$ (red; 
  known to suffer from strong systematic
  effects), or a much shorter (by a factor 6) temporal baseline$^{6}$ (green).
  The third estimate$^{7}$ (blue) was derived assuming that the
  line-of-sight velocity gradient observed in Sculptor is due to
  perspective effects (``apparent rotation'').
  However, in the presence of intrinsic rotation 
  the PM derived in this way will be based on an incorrect assumption.}\label{comparison}
\end{figure}

Our final catalog is shown in Fig.~\ref{vpd}. Very distant
objects such as background galaxies and quasars do not move and thus if
present will have an apparent non-zero proper
motion as a result of our procedure that sets Sculptor at rest$^{15,16}$.  
Although there are no known quasars in our field of view, we were able to identify two
background galaxies using the \gaia astrometric excess noise parameter$^{17}$, 
and confirmed by eye (see Figs.~\ref{fov}b and \ref{fov}c). Even
though these are extended sources, their cores are well fit by a point
source-like PSF, making them reliable for defining the absolute
reference frame. The relative PMs measured for these two galaxies are 
red crosses in Fig.~\ref{vpd}b. The fact that they both lie in the
same region of this PM diagram supports our analysis.  We adopted their
weighted mean relative proper motion (blue cross in Fig.~\ref{vpd}b)
as the zero-point, thus the absolute PM for Sculptor is
($\mu^{abs}_{\alpha}\cos\delta$, $\mu^{abs}_{\delta}$)=($-0.20\pm0.14,
-0.33\pm0.11$)~\masyr, which corresponds to ($-79.6\pm55.7, -131.4\pm43.8$)~\kms~ assuming a distance of $84 \pm 2$~kpc to Sculptor$^{18}$. Fig.~\ref{vpd}c shows that the motions of the
stars in the field are coherent.  Finally,
Fig.~\ref{comparison} compares our PM measurement to previous
estimates\footnote{during the publication process of this paper, a new estimate has been provided by Sohn et al.2017$^{44}$}. More details and a thorough description of the extensive
tests we have performed are reported in the Methods section.

To compute the orbit of Sculptor around the Milky Way and also to
quantify the effect of ``apparent rotation''$^{7}$, we combine our
absolute PM measurement with literature values of the line of sight
velocity$^{4}$ $v_{\rm los}$, distance$^{18}$, and sky position of Sculptor. We
use these as initial conditions (and also consider PMs within
1$\sigma$ of the measured values) for the integration of orbits in a
multi-component Galactic potential$^{19}$. These show that Sculptor
moves on a relatively high inclination orbit and that it is currently
close to its minimum distance to the Milky Way, as we find its peri-
and apocenter radii are $r_{\rm peri} = 73^{+8}_{-4}$~kpc and $r_{\rm
  apo} = 222^{+170}_{-80}$~kpc. The values of these orbital parameters
depend on the assumed mass for the Milky Way halo, but variations
of 30\% lead to estimates within the quoted uncertainties (see the
Methods section for more details).

Finally, we deduce the maximum apparent rotation for this orbit to
be $2.5$~\kmsdeg \linebreak at a position angle $\sim
18$~$\deg$. Therefore if we correct the velocity
gradient along the major axis previously measured$^{4}$ in Sculptor for this apparent rotation, we
find an intrinsic rotation signal along this axis of amplitude
$5.2$~\kmsdeg.  This implies that at its half-light radius, $v_{\rm
  rot}/\sigma_{\rm los} \sim 0.15$, for a line-of-sight velocity
dispersion$^{4}$ $\sigma_{\rm los} = 10$~\kms. Given the
large pericentric distance and the small amount of rotation we have
inferred, this implies that Sculptor did not
originate in a disky dwarf that was tidally perturbed by the
Milky Way$^{20}$.

We determined the internal transverse motions of the stars in
Sculptor using a sub-sample selected such that: ($i$) $18.4< G<19.1$
mag, to avoid stars in the HST non-linear regime and those where the
\gaia positional errors are more uncertain$^{21}$; ($ii$) the
errors on each of the PM components are smaller than $0.07$~\masyr
(corresponding to $27.9$~\kms~ at the distance of Sculptor); ($iii$)
the total PM vector is smaller than 0.23~\masyr (i.e.\ $91.6$~\kms,
this limit is set by the apparent PM of the background galaxies).
There are 15 stars that satisfy these criteria and hence have the best
PM measurements.

\begin{figure}
 \begin{minipage}{0,4\textwidth}
 \centering
 \vskip 0.5 cm
 \includegraphics[scale=0.65, trim=0 0 0 -0.5cm]{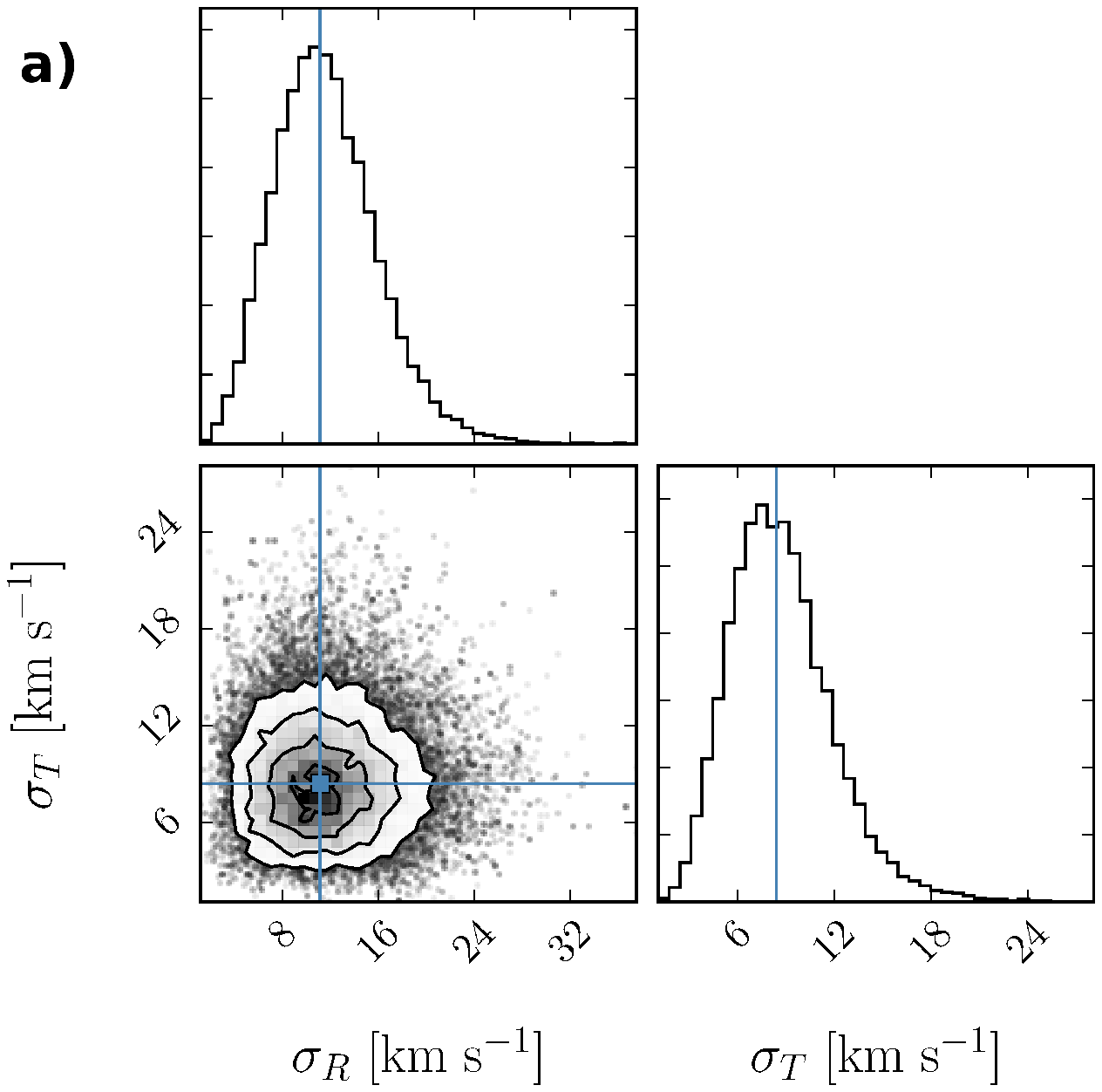}
 \end{minipage}
 \begin{minipage}{0,6\textwidth}
  \centering
  \includegraphics[height=3.5cm, width=6cm]{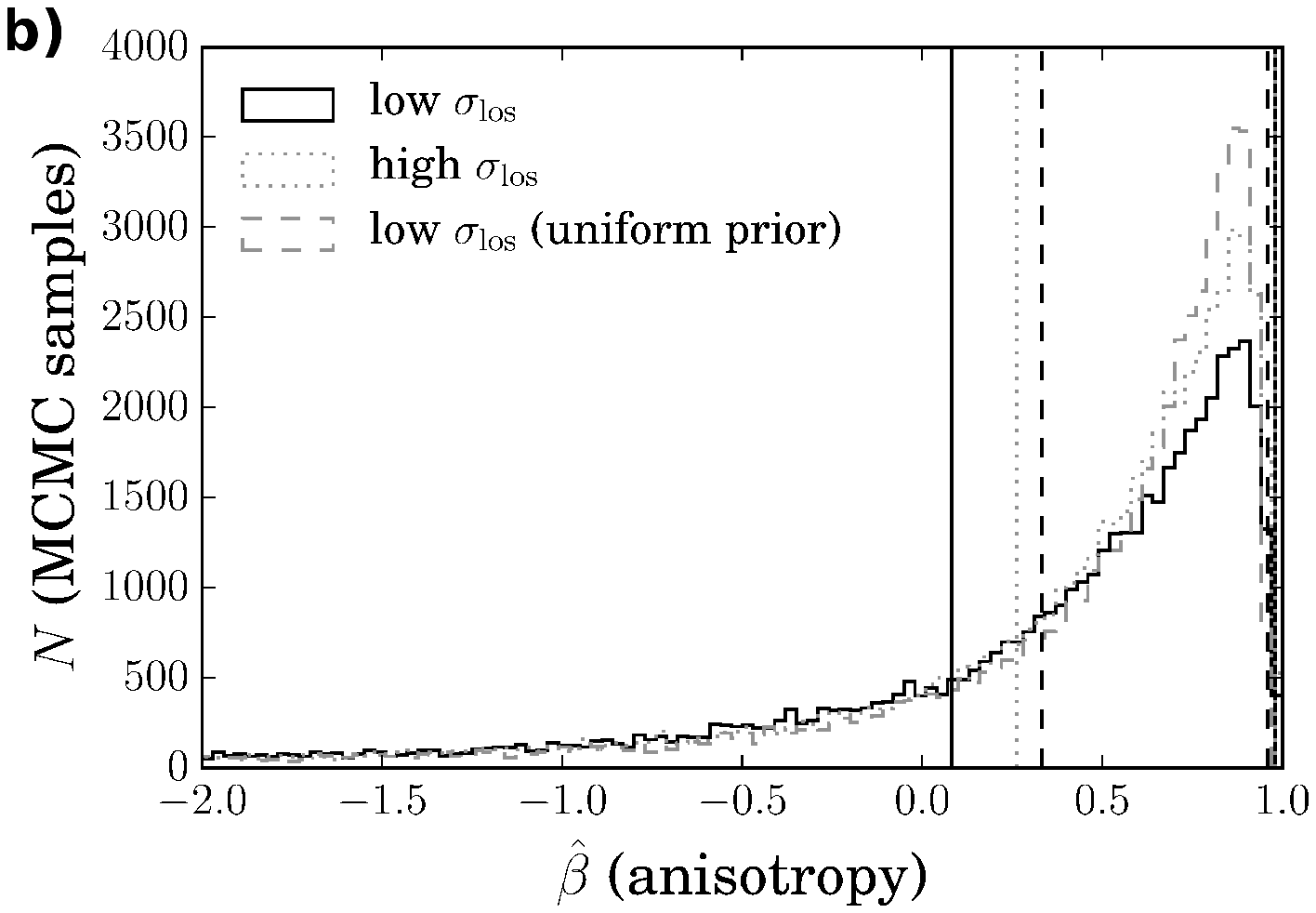}
  \vskip 0.5cm
  \includegraphics[height=3.5cm, width=6cm]{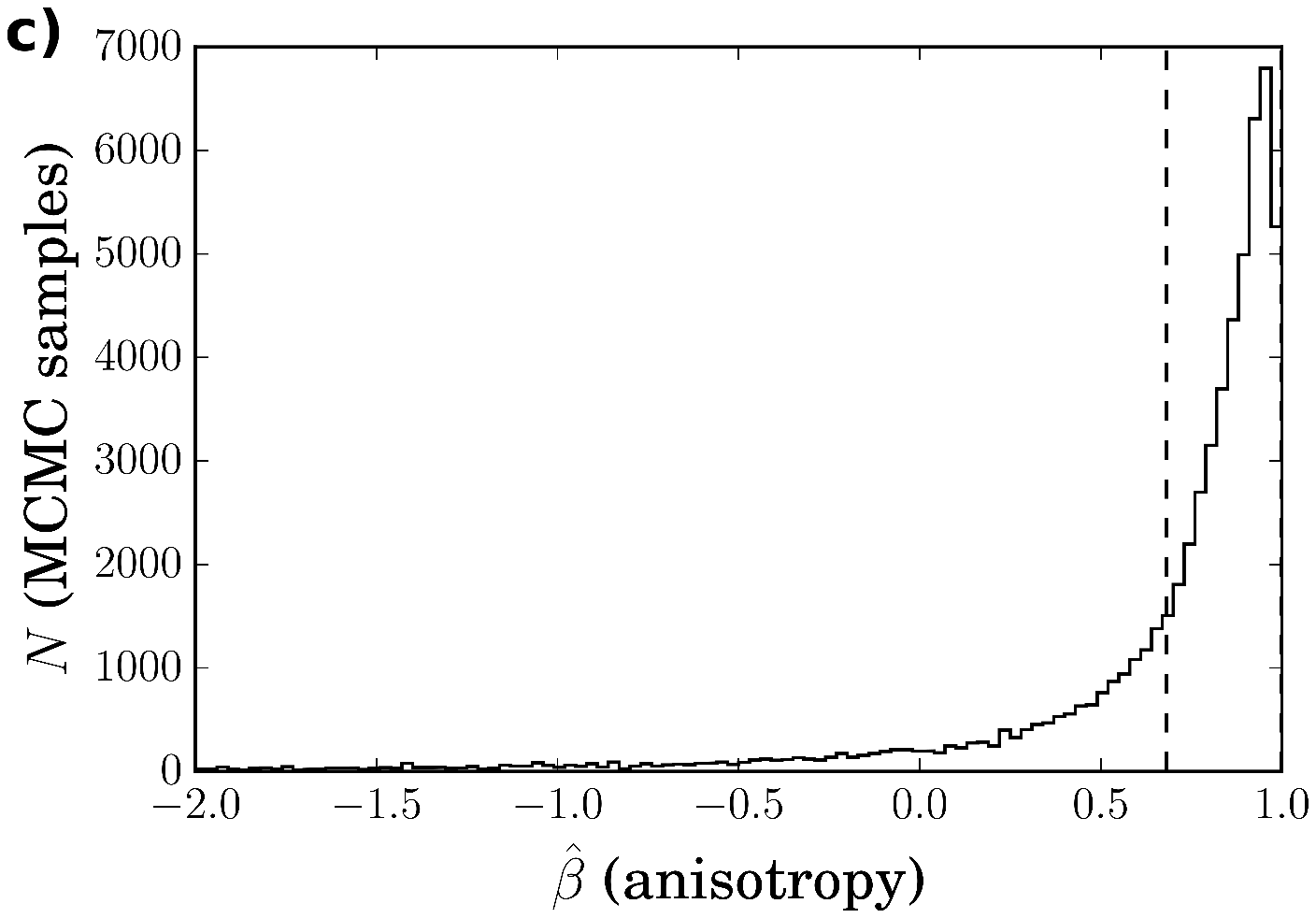}
 \end{minipage}
\caption{{\bf 2D velocity dispersion and orbital anisotropy of
    Sculptor.}  a) shows the posterior probability distribution for
  the projected velocity dispersions $\sigma_{R}$ and $\sigma_{T}$ for
  the sample of 15 stars with the best PM measurements. Their maximum
  a posteriori (MAP) values are indicated with the solid (blue)
  lines. b) shows the resulting distribution of the anisotropy
  parameter $\hat{\beta}$ at a radius $\hat{r} \ge R_{HST}$, where
  $R_{HST} \sim 7.6$\arcmin~ is the average projected distance of
  stars from the center of Sculptor. The solid and dashed histograms
  are computed using $\sigma_{\rm los}$ for these stars (assuming a
  Gaussian and flat priors respectively), and the dotted histogram is
  for a more commonly used value of $\sigma_{\rm los} = 10$~\kms. The
  MAP values for the anisotropy are, for the low $\sigma_{\rm los}$,
  $\hat{\beta}_{\rm MAP} = 0.86_{-0.83}^{+0.12}$ (Gaussian prior),
  $\hat{\beta}_{\rm MAP} = 0.83_{-0.55}^{+0.14}$ (flat prior), and
 $\hat{\beta}_{\rm MAP} = 0.86_{-0.64}^{+0.09}$ for the high
  $\sigma_{\rm los}$. c) shows the posterior probability distribution
  for $\hat{\beta}$ for the metal-rich subsample, using their
  $\sigma_{\rm los}$. The vertical lines in panels b) and c) mark the
  68\% highest posterior density intervals around the MAP
  values.}\label{fig:beta}

\end{figure}

We model the velocity dispersion of this sample using a
multivariate Gaussian. The parameters of this distribution are the
mean velocities in the radial and tangential directions on the plane
of the sky ($v_{0,R}, v_{0,T}$), the dispersions ($\sigma_R,
\sigma_T$) and their correlation coefficient $\rho_{R,T}$. We use
Bayes theorem to derive the posterior distribution for these
parameters (assuming a Gaussian-like prior on the dispersions) from a
Markov Chain Monte Carlo (MCMC) algorithm$^{22}$.  We
find $\sigma_{R} =11.5 \pm 4.3$~\kms~ and $\sigma_{T}=8.5\pm3.2$~\kms,
as shown in Fig.~\ref{fig:beta}a.

If we assume spherical symmetry and neglect rotation (see the Methods
section for details), we can use the Jeans equations to find a
relation$^{23}$ between the velocity dispersions measured at
$R_{HST}$ (the location of our fields) and the value of the anisotropy
$\hat{\beta} = \beta({\hat r})$ where ${\hat r} \ge R_{HST}$:
\begin{equation}
\hat{\beta} = 1 - \frac{\sigma^2_T}{\sigma^2_\text{los} + \sigma^2_{R} - \sigma^2_{T}}.
\end{equation}
We determine $\sigma_\text{los} \sim 6.9$~\kms for 10 stars in common with a spectroscopic catalog$^{24}$. Using the MCMC chain
samples, we obtain the probability distribution for $\hat{\beta}$
shown in Fig.~\ref{fig:beta}b.  The two other histograms in this panel
depict the results obtained assuming a flat-prior (dashed) or the more
often quoted value $\sigma_{\rm los} \sim 10$~\kms (dotted). In
all cases, radial anisotropy is clearly favored, with a median value
$\hat{\beta} \sim 0.46$ and the maximum a posteriori $\hat{\beta}_{\rm
  MAP} \sim 0.86$.

This is the first ever determination of the value of the anisotropy
$\beta$ in an external galaxy. The
anisotropy is the key missing ingredient to robustly establish the distribution
of matter in Sculptor, reflected in a longstanding unresolved
debate$^{25,26,27,10}$, as to whether or not this galaxy
has the cuspy profile$^{28}$ predicted by the
concordance cosmological model in which dark matter is cold, 
constituted by weakly interacting particles$^{2}$.

The value of $\beta$ we have measured is surprising. A
review$^{9}$ of the literature indicates that most
previous works have assumed spherical symmetry and derived, for a
variety of mass models of Sculptor, $\beta \le 0$ for $\beta$ constant with
radius.  However, no physical system can have a constant anisotropy
and $\beta \sim 0.8$ with a light density profile that has a central
slope $\gamma(0) \sim 0$, since $\gamma$ has to satisfy $\gamma \ge 2
\beta$ in the spherically symmetric limit$^{29}$. Therefore,
in this context, our result shows that the anisotropy in Sculptor
cannot be constant with radius. Our measurement also rules out the
simplest predictions for Sculptor's anisotropy based on
the alternative gravity model known as MOND$^{30}$.

Our results highlight the necessity to go beyond the standard
assumptions. We may need to consider that Sculptor's dark halo may be
axisymmetric or even triaxial. Alternatively and quite plausibly our
measurement may be biased towards the colder, more centrally
concentrated, metal-rich(er) subcomponent of
Sculptor$^{4}$. Of the 15 stars in our best PM sample, 9
have a metallicity measurement$^{24}$ (see Methods section for details) and 6 of these have
[Fe/H]$>-1.4$ dex, indicating that about half could belong to this
subcomponent of Sculptor. From the 11 stars in our sample with
[Fe/H]$>-1.4$ dex, $18.4 \le G \le 21$, and that satisfy also the
quality criteria, we determine the anisotropy to be clearly radial
with $\hat{\beta}^{MR}_{\rm MAP} = 0.95_{-0.27}^{+0.04}$ and a median
$\hat{\beta}^{MR} = 0.82$ at a distance $\hat{r} \ge R_{HST}$, as
shown in Fig.~4c. This value is in excellent
agreement$^{10}$ with predictions if Sculptor's metal-rich component is embedded 
in a cuspy dark halo profile. It remains
to be seen if such a high value can also be consistent with cored models, since
those published$^{4,26}$ typically predict
lower, though still radial, anisotropy.  Another intriguing
question is what formation mechanism produces a population of stars
moving on such very elongated orbits.

{\bf Acknowledgements:} We have made use of data from the European Space Agency mission
\gaia ({\tt http://www.cosmos.esa.int/gaia}), processed by the \gaia
Data Processing and Analysis Consortium (DPAC, {\tt
  http://www.cosmos.esa.int/web/gaia/dpac/consortium}). Funding for
DPAC has been provided by national institutions, in particular the
institutions participating in the \gaia Multilateral Agreement.  This
work is also based on observations made with the NASA/ESA Hubble Space
Telescope, obtained from the Data Archive at the Space Telescope
Science Institute, which is operated by the Association of
Universities for Research in Astronomy, Inc., under NASA contract NAS
5-26555. A.H. and L.P. acknowledge financial support from a Vici grant
from the Netherlands Organisation for Scientific Research. M.B. and
A.H. are grateful to NOVA for financial support.

{\bf Authors' contributions:} D.M. performed the data analysis and the proper motion measurements,
 M.B. developed the statistical tools, A.H. derived the relations between observables and orbital anisotropy, coordinated the work
 and led the scientific interpretation, L.P. performed the orbit computation, A.B. and E.T. contributed
 to the presentation of the paper. All the authors critically contributed to the work presented here.
 Correspondence and requests for materials should be addressed to massari\at astro.rug.nl”

\title{{\bf Methods Section}}

\section{Description of the HST data and procedures}

To measure the proper motions (PMs) of stars in the Sculptor dwarf spheroidal galaxy we
used two epochs of data obtained with the two best astrometric space
facilities available at the moment: the HST and the \gaia mission.
The first epoch of observations was acquired with the Wide Field
Channel (WFC) of the Advanced Camera for Survey (ACS) on board the
HST. This camera is made up of two $2048 \times 4096$ pixel detectors
separated by a gap of about 50 pixels. Its pixel scale is $\sim0.05
\arcsec\,$pixel$^{-1}$, for a total field of view (FoV)
$\sim200\arcsec \times 200\arcsec$.  The data set (GO-9480, PI:\
Rhodes), consists of two overlapping pointings separated by about 2\arcmin.  In turn, the first pointing is split in
five $400$ sec long exposure images in the F775W filter.  The second
pointing is made up of six exposures with the same
characteristics. The overlapping FoV has thus been observed 11
times. This data set has been acquired on the 26th of September, 2002.

We retrieved from the archive only \_FLC images, which are corrected
for charge transfer efficiency (CTE) losses by the pre-reduction
pipeline adopting a pixel-based correction$^{31,32}$.  
The data-reduction was
performed with the \texttt{img2xym$\_$WFC.09$\times$10} program$^{33}$.  We
treated each chip of each exposure independently, and we obtained a
catalog with positions, instrumental magnitudes and Point Spread
Function (PSF) fitting-quality parameter for each of them. Stellar
positions were corrected for filter-dependent geometric distortions$^{11}$.  
We then cross-matched the single catalogs to compute 3$\sigma$-clipped average positions,
magnitudes and corresponding uncertainties (defined as the rms of the
residuals around the mean value).  We finally built the total HST
catalog after excluding all the saturated sources and those that were
measured less than 4 times.

\section{Error analysis}\label{errors}

Since for this study a good control of all the uncertainties is fundamental,
in the following we summarize every source of measurement error in an attempt
to find and correct possible unaccounted for terms.

\subsection{Intrinsic errors}

\subsubsection{HST}

\begin{figure}
 \centering%
\includegraphics[scale=0.5]{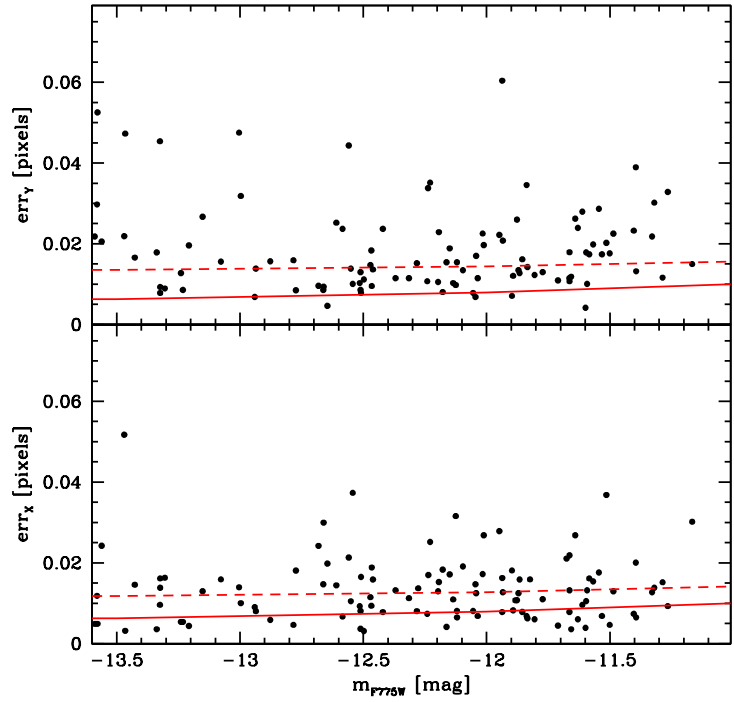}
\caption{{\bf HST internal errors.} Comparison between our estimated single-epoch positional
  errors and models$^{34}$ given by the solid red
  lines. Our estimate of the errors on the data are systematically
  above the prediction, and the addition to the model of a 0.01 pixel
  term (dashed red lines) corresponding to the typical residuals in the
  geometric distortion solution for the F775W filter leads to better 
  agreement.}\label{hsterr}
\end{figure}

From the analysis of many HST dithered images, a general trend for the
behavior of ACS/WFC single exposures positional errors as a function
of instrumental magnitude and adopted filter has been
derived$^{34}$. This trend has been modeled for three filters (F435W,
F606W and F814W), but very similar results were found for all of them,
and especially for the two redder ones.  Our exposures have been
observed in the filter F775W (instrumental magnitudes were
  calibrated onto the VEGAmag system using publicly available$^{35}$ aperture corrections and
  zeropoints), so that it is reasonable to
compare the positional errors we obtained with the model describing
F606W and F814W.  Such a comparison is shown in Fig.~\ref{hsterr}. To
compute our single-exposure positional errors, we multiplied the rms
values in the HST global catalog obtained as described above, by
sqrt($N$), where $N$ is the number of times each star has been
measured.

Single-exposure errors computed in this way still contain another
source of uncertainty given by possible residuals in the geometric
distortion solution.  It has been reported$^{36}$ that the distortion
solution for the F775W filter is slightly worse than e.g.\ that for the
F606W filter because of the lower number of images available used for
modeling. The expected residuals should be of the order of $0.01$
pixels$^{36}$. Indeed, this explains very well why our
errors are located systematically above the expectation given by the red solid
line in Fig.~\ref{hsterr}. By adding in quadrature an additional term of $\sim 0.01$
pixels, which mimics the effect of distortion residuals, the expected
trend (dashed red line) matches well the median behavior obtained
from our data. Therefore, we conclude that the estimated errors for
the HST first-epoch position are reasonable and robust. 

\subsubsection{\gaia}
The \gaia positional uncertainties and correlations have been extensively
analyzed and discussed in the recent literature$^{17,37}$. Their determination will
certainly improve in the next data releases, but they are currently in
the best shape allowed by the amount of data collected so far. We
therefore take the errors at their face value.

\subsection{Systematic uncertainties}

PM measurements can be affected by several systematic uncertainties. In the following
we test our measurements against a comprehensive list of systematic effects, based on the
prescriptions described in previous work$^{34}$.

\begin{figure}
 \centering%
\includegraphics[scale=0.5]{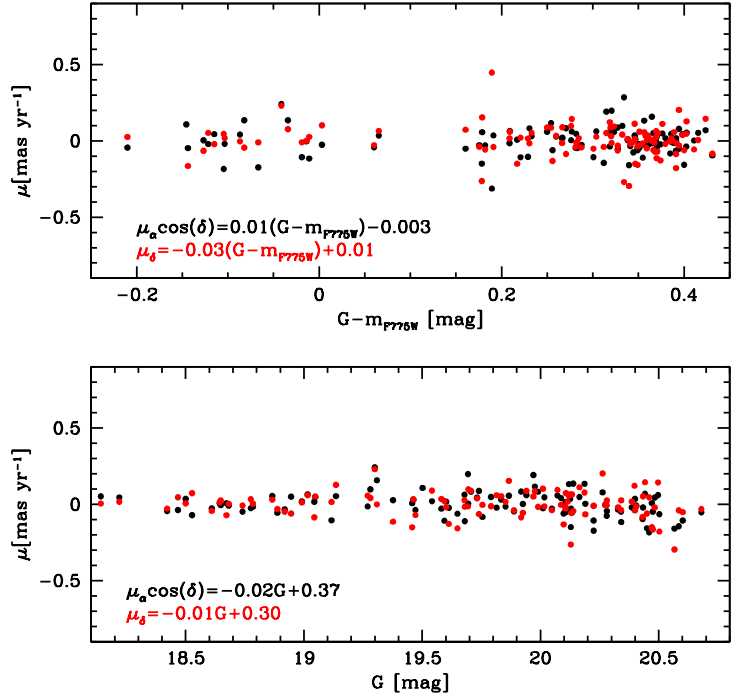}
\caption{{\bf Systematic trends with respect to photometry.}{\it Top panel:} PMs versus observed (G-m$_{F775W}$) color. None of the two PM components show
any systematic trend. The best linear fit parameters are quoted in the lower-left corner.
{\it Bottom panel:} same but for the PMs versus \gaia G-band magnitudes.}\label{photo}
\end{figure}

{\it Chromatic effects}. Differential chromatic refraction$^{38}$ (DCR) is one of the most common sources of
systematic uncertainties on astrometric measurements. This is due to the fact that DCR shifts the
position of photons on the detector proportionally to their wavelength and to the zenithal angle of the
observations. Since this effect is induced by the atmosphere, our data taken from space facilities
should be unaffected, but possible chromatic effects could still play a role.
We checked for this by looking for trends of our PMs as a function of color (G-m$_{F775W}$).
As evident in the top panel of Fig.~\ref{photo} (where the two PM components $\mu_{\alpha}\cos(\delta)$ and $\mu_{\delta}$
are shown with black and red symbols, respectively) no such trends are apparent. In fact
the best least squares linear fit, $\mu = a_\mu + b_\mu (G-m_{F775W})$, has coefficients that are 
consistent with zero within 1$\sigma$ (e.g $b_{\mu\alpha_*}=0.01\pm0.09$~\masyrmag
and $b_{\mu\delta}=-0.03\pm0.09$~\masyrmag). We can therefore rule out the presence of systematic
chromatic effects affecting our PMs.

\begin{figure}
 \centering%
\includegraphics[scale=0.5]{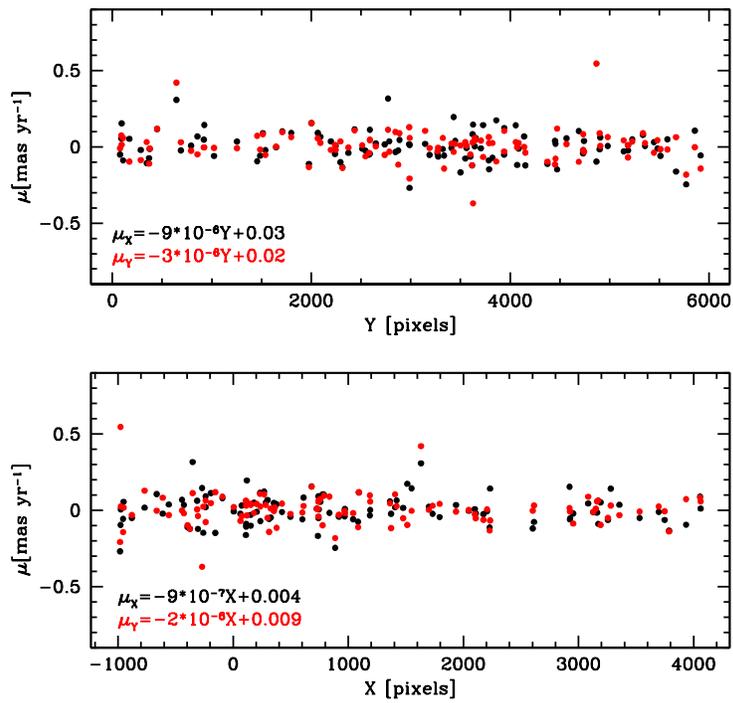}
\caption{{\bf Systematic trends with respect to position.} PM components along the ACS detector X- and Y- directions versus the position on the detector.
All the best linear fits are consistent with no systematic trend with location.}\label{pos}
\end{figure}

{\it CTE losses}. Defects in the silicon lattice of the ACS detector can lead to
an inefficient read-out of the charge that causes deferred-charge trails developing from each source along
the vertical direction$^{32}$. This effects tends to systematically move the centroid of sources
in the same (vertical) direction, and more significantly affects faint objects$^{39}$.
The images we used in this study have already been corrected for CTE losses, but we further checked for
the existence of possible residuals by looking for trends among our measured PMs and magnitude (faint
stars should be more affected) and positions (trends along the ACS Y-direction should be observed).
The first of these tests is shown in the bottom panel of Fig.~\ref{photo}, where the two PM components are
plotted against \gaia G-band magnitudes. As in the previous case, no trend is found (the slopes being
$b_{\mu\alpha*}=-0.02\pm0.08$~\masyrmag and $b_{\mu\delta}=-0.01\pm0.09$~\masyrmag).
The second test is shown in Fig.~\ref{pos}. We rotated the PMs by $24.75$ degrees, such that their X- and Y-
components correspond to the horizontal and vertical direction of the ACS detector. Again, in all cases
the slopes of the best linear fit are fully consistent with zero, i.e. no trends are apparent. We can then conclude that
residuals CTE effects are not affecting our measurements.

{\it Other systematic effects.} In general, other not well identified
systematic effects could affect our PM measurements or their estimated
uncertainties. We checked for their presence by looking for trends
between PMs and all the other measured quantities (photometric
parameters, positions, quality of the PSF fitting, astrometric excess noise), finding none.  Therefore, after
this analysis we can conclude that our PM measurements do not suffer
from (the better) known systematic effects.

Possible global systemic motions of the dSph like
expansion/contraction or rotation on the plane of the sky could
translate into systematic uncertainties on our absolute PM
estimate. However, given the large distance of Sculptor (we adopt
throughout the paper a distance of $84$ kpc, obtained from the
analysis of RR Lyrae variable stars$^{18}$), they are
negligible compared to the uncertainty on the absolute zero-point.
For example, if we assume that the total rotational signal of $7.6$~\kmsdeg~ reported in the literature$^4$ corresponds to rotation on the plane of the
sky, then the corresponding PM at the location of our HST FoV would be
of only $0.003$~\masyr.

\section{The orbit of Sculptor around the Milky Way and its apparent rotation}

We use the observed position on the sky, distance, heliocentric radial
velocity and our newly obtained PM measurements of Sculptor to derive
its orbit. In a right-handed Cartesian heliocentric reference frame, where $X$ points towards the Galactic center, $Y$ in the
direction of rotation and $Z$ is positive towards the Galactic North
pole, Sculptor lies at $(X,Y,Z) = (3, -9.5, -83.4)$~kpc and moves with
velocity $(V_X,V_Y,V_Z)=(143.3, -76, -90.3)$ \kms. We then correct for the Sun's
position and velocity w.r.t. the Galactic center$^{40}$ assuming
$(X_\odot,Y_\odot,Z_\odot) = (-8.3, 0, 0.014)$~kpc, and
$(V_{X,\odot},V_{Y,\odot},V_{Z,\odot}) = (11.1, 240.24,
7.25)$~\kms.  We integrate these initial conditions, together with 100
random realizations assuming that the errors in the observables are
Gaussian, in an axisymmetric Galactic potential for 4 Gyr forward and
backward in time using an 8$^{th}$ order Runge-Kutta method.  The
Galactic potential$^{19}$ has several components: a flattened bulge, a
gaseous exponential disc, thin and thick stellar exponential discs and
a flattened ($q=0.8$) dark matter halo.  The total baryonic (stars and
cold gas) mass of the model is $M_{\rm bary} = 5.3\times
10^{10}\msol$, while the dark halo follows an NFW$^{28}$ profile whose
virial mass is $M_{200}=1.3\times 10^{12}\msol$ and its concentration
$c_{200} = 20$. As reported in the main part of the paper, we find
that Sculptor has recently (approximately $170$ Myr ago) reached its minimum
distance to the Milky Way, and is currently moving outwards. The peri-
and apocenter radii are $r_{\rm peri} = 73^{+8}_{-4}$~kpc and $r_{\rm
  apo} = 222^{+170}_{-80}$~kpc, and the orbit has a relatively high
inclination of $88$ deg. These values are, of course
dependent on the characteristic parameters of the Galactic
potential. To give a flavor of how they change we vary the mass of the Milky Way halo by 30\%. We find
that for $M_{200}=0.9\times 10^{12}\msol$, then $r_{\rm peri} =
83^{+2}_{-10}$~kpc and $r_{\rm apo} = 475^{+210}_{-175}$~kpc, while for
$M_{200}=1.7\times 10^{12}\msol$, $r_{\rm peri} = 73^{+3}_{-2}$~kpc and
$r_{\rm apo} = 143^{+34}_{-21}$~kpc. As expected, only the apocentric
distance varies strongly with $M_{200}$. In fact, if we assume $M_{200}$ to be half of our fiducial
value (i.e. $0.65\times 10^{12}\msol$), then Sculptor would be unbound.

\begin{figure}
\includegraphics[width=\columnwidth]{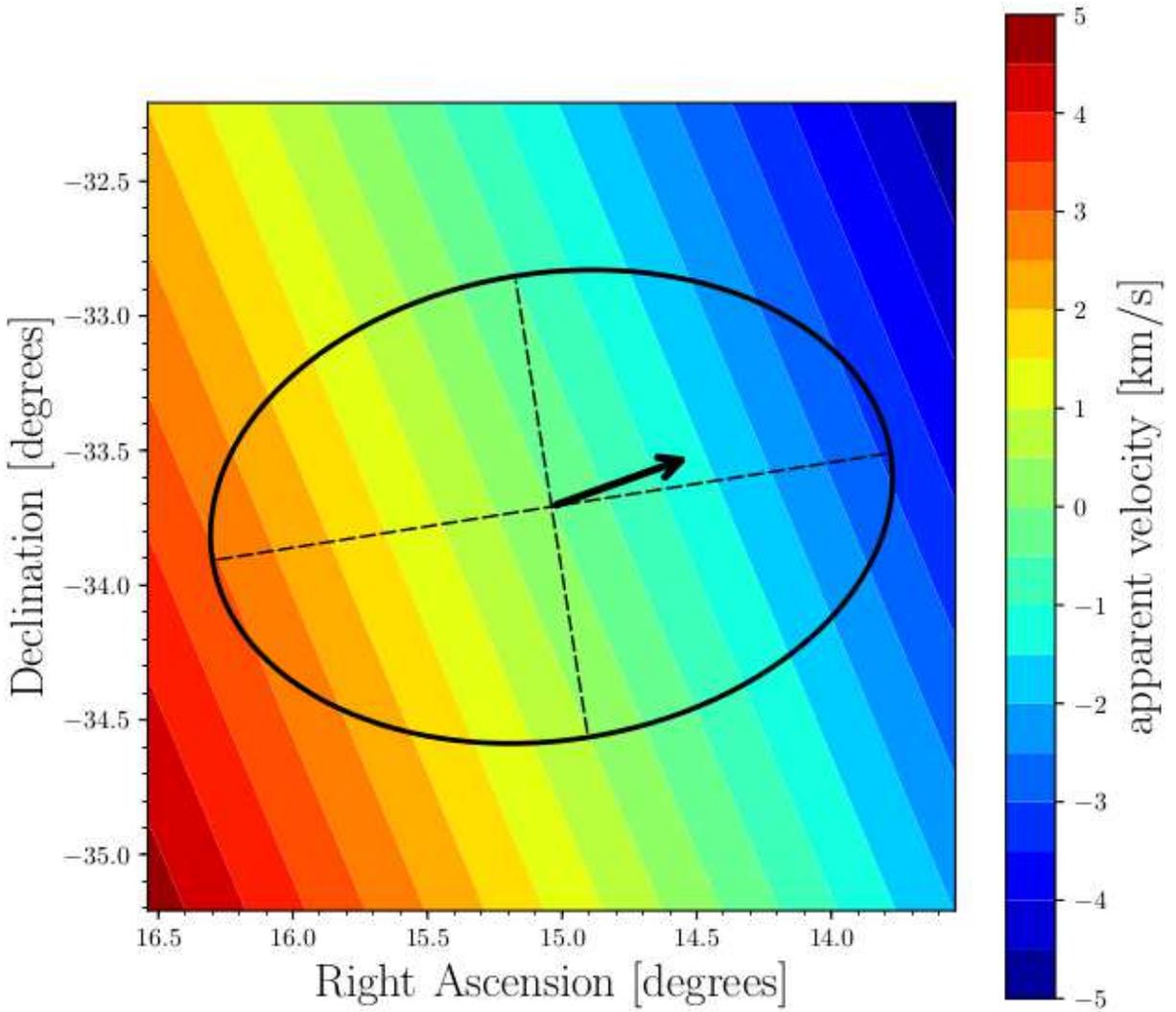}
\caption{{\bf Apparent velocity field induced by the orbital motion of
  Sculptor}. The black ellipse corresponds to the fiducial tidal radius$^{13}$,
  the black dashed lines indicate the major
  and minor axis of the galaxy, while the black arrow gives the
  projected direction of motion. The color coding
  represents the apparent velocity field with respect to the Sun in steps of $0.5$
  \kms.}
  \label{fig:appvrot}
\end{figure}

Now that we have determined the orbital motion of Sculptor, we may
quantify the magnitude of the ``apparent'' rotation. The total
apparent velocity field induced by the orbit is shown in
Fig.~\ref{fig:appvrot}, where the black ellipse corresponds to the
tidal radius of Sculptor$^{13}$ and the direction of
its PM is indicated by the black arrow. The velocity field, color-coded in steps of
$0.5$ \kms, has a maximum magnitude of $2.5$ \kmsdeg at $\rm{PA}\simeq
18\deg$, that is projected to an apparent velocity gradient of $2.4$
\kmsdeg~ along the major axis and $0.7$~\kmsdeg~ along the minor axis.

\section{Velocity dispersion and anisotropy}\label{aniso}

In this section we describe the procedure for deriving the velocity
dispersion on the plane of the sky as well as the velocity anisotropy of Sculptor. 

We transform the PM components from the equatorial reference to radial and tangential components on the
plane of the sky according to the equatorial-polar coordinates relation$^{41}$:
$$\left[ \begin{array}{c} \mu_{R} \\ \mu_{T} \end{array} \right] = \begin{bmatrix} \cos(\phi) & \sin(\phi) \\ -\sin(\phi) & \cos(\phi) \end{bmatrix} \times \left[ \begin{array}{c} \mu_{\alpha}\cos(\delta) \\ \mu_{\delta} \end{array} \right],$$
where $\phi=\arctan(y/x)$, and $x$ and $y$ are the (local Cartesian)
gnomonic projected coordinates. Uncertainties are fully propagated
taking into account the correlation coefficient between \gaia's RA and
DEC estimates. The projected velocities in the radial and tangential
direction therefore are $v_{R,T} = 4.74 \mu_{R,T} d$, with $d$ the
distance to Sculptor.

We model the velocity dispersion for the sample selected as described
in the main body of the paper, by a multivariate Gaussian including a
covariance term. This Gaussian is characterized by velocity
dispersions in the (projected) radial and tangential directions
($\sigma_R, \sigma_T$), their correlation coefficient $\rho_{R,T}$ and
the mean velocities ($v_{0,R}, v_{0,T}$). The posterior for these
parameters ${\bf p} = (\sigma_R, \sigma_T, v_{0,R}, v_{0,T},
\rho_{R,T})$, including the data $D$, is given by Bayes theorem:
\begin{equation}
	p(\sigma_R, \sigma_T, v_{0,R}, v_{0,T}, \rho_{R,T}|D) =p(D|\sigma_R, \sigma_T, v_{0,R}, v_{0,T}, \rho_{R,T}) p(\sigma_R, \sigma_T, v_{0,R}, v_{0,T}, \rho_{R,T}) / p(D).\\
\label{eq:bayes}
\end{equation}
The likelihood here, $p(D|{\bf p})$, is a product of Gaussians,
$\mathcal{N}$, and the covariance matrix is the sum of the covariance
matrices associated to the intrinsic kinematics of the population and
to the measurement uncertainties (which is equivalent to a convolution
of the two Gaussians representing these contributions), i.e.
\begin{eqnarray}
	p(D|\sigma_R, \sigma_T, v_{0,R}, v_{0,T}, \rho_{R,T}) \!\!\!\!&=& \!\!\!\!\prod_i  \mathcal{N}\bigg(\begin{bmatrix} v_{R,i} \\ v_{T,i}  \end{bmatrix}, \begin{bmatrix} v_{0,R} \\ v_{0,T} \end{bmatrix}, \Sigma_i\bigg),\\
	\Sigma_i \!\!\!\!&=& \!\!\!\!\begin{bmatrix} \sigma_R^2 & \rho_{R,T}\sigma_R \sigma_T\\ \rho_{R,T}\sigma_R \sigma_T & \sigma_T^2 \end{bmatrix} + \begin{bmatrix} \epsilon_{\sigma_R,i}^2 & \rho_{R_i,T_i}\epsilon_{\sigma_R,i} \epsilon_{\sigma_T,i} \\ \rho_{R_i,T_i}\epsilon_{\sigma_R,i} \epsilon_{\sigma_T,i}  & \epsilon_{\sigma_T,i}^2. \end{bmatrix}
\end{eqnarray}
Furthermore, for the prior $p({\bf p})$ in Eq.~(\ref{eq:bayes}), we
assume it has a weak Gaussian-like form for the correlation coefficient
of the intrinsic kinematics of the population (with mean 0, and
dispersion 0.8) while a flat prior is assumed for the mean velocities.  We explore two different priors
for the velocity dispersion: a Gaussian-like for the (logarithm of the) velocity dispersions
(with mean $\log_{10} 12$[\kms], and unity dispersion), and a uniform prior. 

We have used a Markov Chain Monte Carlo (MCMC) algorithm$^{22}$ to
estimate the posterior all the parameters, but, except for
$\sigma_{R}$ and $\sigma_{T}$, we consider all as nuisance parameters.
As reported in the main body of the paper, we find for our best PM
sample $\sigma_{R} =11.5\pm 4.3$~\kms~ and
$\sigma_{T}=8.5\pm3.2$~\kms~ in the case of the Gaussian prior (similar values
are obtained for the flat case).

In our analysis we have left the mean projected velocities $v_{0,R}$
and $v_{0,T}$ as free (nuisance) parameters, and find values entirely
consistent with those determined in the main body of the paper. If
Sculptor would rotate with an amplitude of $5.2$~\kmsdeg, 
this would induce a gradient in the field where
our stars are found of order 0.5~\kms, that therefore would be
negligible.

The anisotropy $\beta(r)$ provides a measure of the intrinsic orbital
distribution of the system, and is defined as $\beta = 1 -
\displaystyle\frac{\sigma^2_t}{2\sigma^2_r}$, where $\sigma_t$ and
$\sigma_r$ are the intrinsic (3D) velocity dispersions in the
tangential and radial directions, respectively. To obtain an estimate of the orbital
anisotropy $\beta$ from the observables $\sigma_{\rm los}$, $\sigma_R$
and $\sigma_T$ we use the spherical Jeans equations. These link the
measured dispersions with intrinsic properties of the system, namely
$\sigma_r(r)$, $\beta(r)$, and the light density profile ($\nu_*(r)$ in 3D
and projected $I_*(R)$) as follows$^{23}$:
\begin{eqnarray}
\sigma_{\rm los}^2(R) & = &  \frac{2}{I_\star(R)} \int_{R}^{\infty} \left ( 1 - \beta \frac{R^{2}}{r^2} \right )
\frac{\nu_{\star} \sigma_{r}^{2} r dr}{\sqrt{r^2-R^2}} \,, \label{eq:LOSdispersion} \\
\sigma_R^2(R) & =  & \frac{2}{I_\star(R)} \int_{R}^{\infty} \left ( 1
- \beta+ \beta \frac{R^{2}}{r^2} \right )
\frac{\nu_{\star} \sigma_{r}^{2} r dr}{\sqrt{r^2-R^2}} \,, \label{eq:Rdispersion}\\
\sigma_T^2(R) &  = & \frac{2}{I_\star(R)} \int_{R}^{\infty} \left ( 1 - \beta \right )
\frac{\nu_{\star} \sigma_{r}^{2} r dr}{\sqrt{r^2-R^2}}  \label{eq:phidispersion}. 
\end{eqnarray}
If we define $Q(r) = \nu_{\star} \sigma_{r}^{2} r/\sqrt{r^2-R^2}$, and 
\begin{equation}
f_1(R) = \int_{R}^{\infty} Q(r) dr, \qquad
f_2(R) = \int_{R}^{\infty} \beta(r) \frac{R^2}{r^2} Q(r) dr, \qquad
f_3(R) = \int_{R}^{\infty} \beta(r) Q(r) dr, \nonumber
\end{equation}
then 
\begin{eqnarray}
\sigma_{\rm los}^2(R) & = &  \frac{2}{I_\star(R)} (f_1(R) - f_2(R)), \nonumber \\
\sigma_R^2(R) & =  & \frac{2}{I_\star(R)} (f_1(R) - f_3(R) + f_2(R)), \nonumber \\
\sigma_T^2(R) &  = & \frac{2}{I_\star(R)} (f_1(R) - f_3(R)). \nonumber
\end{eqnarray}
If we do not make any assumptions on $\beta(r)$, we may
use the mean value theorem in the form
\begin{equation}
\int_a^b f(x) g(x) dx = f(c) \int_a^b g(x) dx, \qquad c\in [a,b], \nonumber
\end{equation}
which holds provided $g(x)$ does not change sign in $[a,b]$. In our case we could apply this theorem to say that 
$\exists \, \hat{r} \in [R_{HST},\infty)$ such that $f_3(R_{HST}) = \hat{\beta} f_1(R_{HST})$, where $R_{HST}$ is the location where we have measured the velocity dispersions $\sigma_R$ and $\sigma_T$ with our dataset. This means that
\begin{equation}
 \hat{\beta} = \beta(\hat{r}) = 1 - \frac{\sigma^2_T}{\sigma^2_\text{\rm los} + \sigma^2_{R} - \sigma^2_{T}}, \qquad \qquad {\rm with} \;  \hat{r} \in [R_{HST},r_{\rm max}),
\label{eq:beta_obs}
\end{equation}
where we have used that in reality, Sculptor has a maximum (finite)
radial extent which we denote by $r_{\rm max}$. Note that if $\beta$
is constant, then Eq.~(\ref{eq:beta_obs}) holds at every radius.

As discussed in the main body of the paper, there are indications that
our sample may be dominated by metal-rich stars, a component known to
have its own characteristic spatial distribution and
kinematics$^{42}$. To derive the metallicity of our stars, we took the measured iron spectral index $\Sigma$Fe reported in the spectroscopic sample observed with the MIKE spectrograph
at the Magellan 6.5m telescope$^{24}$, and applied the following relation$^{43}$
\begin{equation}
{\rm [ Fe/H]}=(7.02\pm2.10)\Sigma{\rm Fe} -3.97\pm2.03
\end{equation}
to calibrate it to [Fe/H]. Since the metal-rich and metal-poor
populations of Sculptor have been clearly separated on the basis of
their metallicity$^4$, we preferred to work with the iron spectral index rather than with the mean reduced Mg
index used in other works$^{25,10}$.

To explore further the possibility that our measurement of the
anisotropy could be affected by the presence of the different populations
in Sculptor, we repeat the procedure outlined above to determine the value
of $\hat{\beta}^{MR} = \beta^{MR}(\hat{r}^{MR})$ where again
$\hat{r}^{MR} \in [R_{HST},r^{MR}_{\rm max})$, now using sample that
includes only stars with [Fe/H]$ \ge -1.4$~dex. To enlarge the
statistics, we also include 5 fainter, similarly metal-rich stars. For
this sample we find $\hat{\beta}^{MR}_{\rm MAP} =
0.95_{-0.27}^{+0.04}$, a value that is much more tightly constrained
that the $\hat{\beta}_{\rm MAP}$ obtained using the best PM sample
without a metallicity cut (compare Fig.~4b and 4c in the main body of
the paper). The reason for this is not a decrease in the errors (the
data satisfy the same quality criteria), nor different numbers of
objects, but the heterogeneity present in our original best PM
sample. That is, this sample contained stars drawn from the different
components in Sculptor with their own, apparently rather different
orbital structure. Unfortunately our sample of metal-poor stars with
good PM measurements is too small to make a similar analysis and results in an anisotropy that is
relatively unconstrained.



\begin{thebibliography}

\bibitem[Lin \& Faber(1983)]{lin} 1. Lin, D.~N.~C., \& Faber, S.~M.\, Some implications of nonluminous matter in dwarf spheroidal galaxies, Astrophys. J. Lett., 266, 21-25 (1983)

\bibitem[Strigari(2013)]{stri13} 2. Strigari, L.~E.\ Galactic searches for dark matter, Phys. Rep., 531, 1-88 (2013) 

\bibitem[Walker et al.(2007)]{wal07} 3. Walker, M.~G., Mateo, M., Olszewski, E.~W., et al.\, Velocity Dispersion Profiles of Seven Dwarf Spheroidal Galaxies, Astrophys. J. Lett. , 667, 53-56 (2007) 

\bibitem[Battaglia et al.(2008)]{battaglia08} 4. Battaglia, G., Helmi, A., Tolstoy, E., et al.\ The Kinematic Status and Mass Content of the Sculptor Dwarf Spheroidal Galaxy, Astrophys. J. Lett., 681, L13 (2008) 

\bibitem[Schweitzer et al.(1995)]{sch95} 5. Schweitzer, A.~E., Cudworth, K.~M., Majewski, S.~R., \& Suntzeff, N.~B.\ The Absolute Proper Motion and a Membership Survey of the Sculptor Dwarf Spheroidal Galaxy, Astron. J., 110, 2747-2757 (1995)

\bibitem[Piatek et al.(2006)]{piatek06} 6. Piatek, S., Pryor, C., Bristow, P., et al.\ Proper Motions of Dwarf Spheroidal Galaxies from Hubble Space Telescope Imaging. IV. Measurement for Sculptor, Astron. J., 131, 1445-1460 (2006)

\bibitem[Walker et al.(2008)]{walker08} 7. Walker, M.~G., Mateo, M., \& Olszewski, E.~W.\ Systemic Proper Motions of Milky Way Satellites from Stellar Redshifts: The Carina, Fornax, Sculptor, and Sextans Dwarf Spheroidals, Astrophys. J. Lett., 688, L75 (2008)

\bibitem[Gaia Collaboration et al.(2016a)]{prusti} 8. Gaia Collaboration, Prusti, T., de Bruijne, J.~H.~J., et al.\ The Gaia mission, Astron. Astrophys., 595, A1 (2016) 

\bibitem[Battaglia et al.(2013)]{battaglia-rv} 9. as reviewed in Battaglia, G., Helmi, A., \& Breddels, M.\ Internal kinematics and dynamical models of dwarf spheroidal galaxies around the Milky Way, New Astron. Rev., 57, 52-79 (2013)

\bibitem[Strigari et al.(2017)]{stri17} 10. Strigari, L.~E., Frenk, C.~S., \& White, S.~D.~M.\, Dynamical Models for the Sculptor Dwarf Spheroidal in a ΛCDM Universe, Astrophys. J., 838, 123-132 (2017) 

\bibitem[Anderson(2007)]{jayacs} 11. Anderson, J.\ Variation of the Distortion Solution, Inst. Sci. Rep. ACS 2007-08, 12 pages, 8 (2007) 

\bibitem[Gaia Collaboration et al.(2016b)]{brown17} 12. Gaia Collaboration, Brown, A.~G.~A., Vallenari, A., et al.\ Gaia Data Release 1. Summary of the astrometric, photometric, and survey properties, Astron. Astrophys., 595, A2 (2016) 

\bibitem[Irwin \& Hatzidimitriou(1995)]{irwin95} 13. Irwin, M., \& Hatzidimitriou, D.\ Structural parameters for the Galactic dwarf spheroidals, Mon. Not. R. Astron. Soc., 277, 1354-1378 (1995)

\bibitem[Massari et al.(2017)]{massari17} 14. Massari, D., Posti, L., Helmi, A., Fiorentino, G., \& Tolstoy, E.\ The power of teaming up HST and Gaia: the first proper motion measurement of the distant cluster NGC 2419, Astron. Astrophys., 598, L9 (2017) 

\bibitem[Dinescu et al.(1999)]{dinescu99} 15. Dinescu, D.~I., Girard, T.~M., \& van Altena, W.~F.\ Space Velocities of Globular Clusters. III. Cluster Orbits and Halo Substructure, Astron. J., 117, 1792-1815 (1999)

\bibitem[Sohn et al.(2012)]{sohn12} 16. Sohn, S.~T., Anderson, J., \& van der Marel, R.~P.\ The M31 Velocity Vector. I. Hubble Space Telescope Proper-motion Measurements, Astrophys. J., 753, 7 (2012)

\bibitem[Lindegren et al.(2016)]{lindegren16} 17. Lindegren, L., Lammers, U., Bastian, U., et al.\ Gaia Data Release 1. Astrometry: one billion positions, two million proper motions and parallaxes, Astron. Astrophys., 595, A4 (2016)

\bibitem[Mart{\'{\i}}nez-V{\'a}zquez et al.(2015)]{marvaz15} 18. Mart{\'{\i}}nez-V{\'a}zquez, C.~E., Monelli, M., Bono, G., et al.\ Variable stars in Local Group Galaxies - I. Tracing the early chemical enrichment and radial gradients in the Sculptor dSph with RR Lyrae stars, Mon. Not. R. Astron. Soc., 454, 1509-1516 (2015)

\bibitem[Piffl et al.(2014)]{Piffl2014} 19. Piffl, T., Binney, J., McMillan, P.~J., et al.\ Constraining the Galaxy's dark halo with RAVE stars, Mon. Not. R. Astron. Soc., 445, 3133-3151 (2014)

\bibitem[Mayer et al.(2001)]{mayer2001} 20. Mayer, L., Governato, F., Colpi, M., et al.\ The Metamorphosis of Tidally Stirred Dwarf Galaxies, Astrophys. J., 559, 754-784 (2001) 

\bibitem[Michalik et al.(2015)]{michalik15} 21. Michalik, D., Lindegren, L., Hobbs, D., \& Butkevich, A.~G.\ Gaia astrometry for stars with too few observations. A Bayesian approach, Astron. Astrophys., 583, A68 (2015)

\bibitem[Foreman-Mackey et al.(2013)]{foremanmackey13} 22. Foreman-Mackey, D., Hogg, D.~W., Lang, D., \& Goodman, J.\ emcee: The MCMC Hammer, Public. Astron. Soc. Pac., 125, 306 (2013)

\bibitem[Strigari et al.(2007)]{stri07} 23. Strigari, L.~E., Bullock, J.~S., \& Kaplinghat, M.\ Determining the Nature of Dark Matter with Astrometry, Astrophys. J. Lett., 657, 1-4 (2007)

\bibitem[Walker et al.(2009)]{walker} 24. Walker, M.~G., Mateo, M., \& Olszewski, E.~W.\ Stellar Velocities in the Carina, Fornax, Sculptor, and Sextans dSph Galaxies: Data From the Magellan/MMFS Survey, Astron. J., 137, 3100-3108 (2009) 

\bibitem[Walker \& Pe{\~n}arrubia(2011)]{wal11} 25. Walker, M.~G., \& Pe{\~n}arrubia, J.\ A Method for Measuring (Slopes of) the Mass Profiles of Dwarf Spheroidal Galaxies, Astrophys. J., 742, 20 (2011) 

\bibitem[Amorisco \& Evans(2012)]{amoris} 26. Amorisco, N.~C., \& Evans, N.~W.\, Dark matter cores and cusps: the case of multiple stellar populations in dwarf spheroidals, Mon. Not. R. Astron. Soc., 419, 184-196 (2012)

\bibitem[Breddels et al.(2013)]{bredd} 27. Breddels, M.~A., \& Helmi, A.\ Model comparison of the dark matter profiles of Fornax, Sculptor, Carina and Sextans, Astron. Astrophys., 558, A35 (2013) 

\bibitem[Navarro et al.(1996)]{nfw96} 28. Navarro, J.~F., Frenk, C.~S., \& White, S.~D.~M.\ The Structure of Cold Dark Matter Halos, Astrophys. J., 462, 563-575 (1996)

\bibitem[An \& Evans(2006)]{an06} 29. An, J.~H., \& Evans, N.~W.\, A Cusp Slope-Central Anisotropy Theorem, Astrophys. J., 642, 752-758 (2006) 

\bibitem[Angus(2008)]{angus08} 30. Angus, G.~W.\, Dwarf spheroidals in MOND, Mon. Not. R. Astron. Soc., 387, 1481-1488 (2008)

\end{thebibliography}

\begin{thebibliography}

\bibitem[Anderson \& Bedin(2010)]{jay10} 31. Anderson, J., \& Bedin, L.~R.\ An Empirical Pixel-Based Correction for Imperfect CTE. I. HST’s Advanced Camera for Surveys, Public. Astron. Soc. Pac., 122, 1035-1064 (2010)

\bibitem[Ubeda \& Anderson(2012)]{ubedajay} 32. Ubeda, L., Anderson, J.,\ Study of the evolution of the ACS/WFC charge transfer efficiency,\ STScI Inst. Sci. Rep. ACS 2012-03 (2012)

\bibitem[Anderson \& King(2006)]{jayking06} 33. Anderson, J., \& King, I.,\ PSFs, Photometry, and Astronomy for the ACS/WFC,\ STScI Inst. Sci. Rep. ACS 2006-01 (2006)

\bibitem[Bellini et al.(2014)]{bellini14} 34. Bellini, A., Anderson, J., van der Marel, R.~P., et al.\ Hubble Space Telescope Proper Motion (HSTPROMO) Catalogs of Galactic Globular Clusters. I. Sample Selection, Data Reduction, and NGC 7078 Results, Astrophys. J., 797, 115 (2014)

35. {\tt http://www.stsci.edu/hst/acs/analysis/zeropoints.}

\bibitem[Anderson(2006)]{anderson06} 36. Anderson, J.\ Empirical PSFs and Distortion in the WFC Camera, The 2005 HST Calibration Workshop: Hubble After the Transition to Two-Gyro Mode, 11 (2006)

\bibitem[Arenou et al.(2017)]{arenou2017} 37. Arenou, F., Luri, X., Babusiaux, C., et al.\ Gaia Data Release 1. Catalogue validation, Astron. Astrophys., 599, A50 (2017)

\bibitem[Trippe et al.(2010)]{trippe} 38. Trippe, S., Davies, R., Eisenhauer, F., et al.\ High-precision astrometry with MICADO at the European Extremely Large Telescope, Mon. Not. R. Astron. Soc., 402, 1126-1140 (2010)

\bibitem[Chiaberge et al.(2006)]{chiaberge06} 39. Chiaberge, M., Riess, A., Mutchler, M., Sirianni, M., \& Mack, J.\ Mack, J.\ ACS Charge Transfer Efficiency. Results from Internal and External Tests, The 2005 HST Calibration Workshop: Hubble After the Transition to Two-Gyro Mode, 36 (2006)

\bibitem[Sch{\"o}nrich et al.(2010)]{schon10} 40. Sch{\"o}nrich, R., Binney, J., \& Dehnen, W.\ Local kinematics and the local standard of rest, Mon. Not. R. Astron. Soc., 403, 1829-1833 (2010)

\bibitem[Binney \& Tremaine(2008)]{bin} 41. Binney, J., \& Tremaine, S.\ 2008, Galactic Dynamics: Second Edition, by James Binney and Scott Tremaine.~ISBN 978-0-691-13026-2 (HB).~Published by Princeton University Press, Princeton, NJ USA (2008)  

\bibitem[Tolstoy et al.(2004)]{tolstoy04} 42. Tolstoy, E., Irwin, M.~J., Helmi, A., et al.\ Two Distinct Ancient Components in the Sculptor Dwarf Spheroidal Galaxy: First Results from the Dwarf Abundances and Radial Velocities Team, Astrophys. J. Lett., 617, 119-122 (2004)

\bibitem[Zin \& West(1984)]{zinn} 43. Zinn, R., \& West, M.~J.\, The globular cluster system of the galaxy. III - Measurements of radial velocity and metallicity for 60 clusters and a compilation of metallicities for 121 clusters, Astrophys. J. Suppl., 55, 45-66 (1984) 

\bibitem[Sohn et al.(2017)]{sohn17} 44. Sohn, S.~T., Patel, E., Besla, G., et al.\  Space Motions of the Dwarf Spheroidal Galaxies Draco and Sculptor Based on HST Proper Motions with a ∼10 yr Time Baseline, Astrophys. J., 849 (2017)  

\end{thebibliography}
\end{document}